\documentclass[twoside]{article}
\usepackage{fleqn,espcrc2,amsfonts,amssymb}

\def\D{{\rm d}}

\title{Critical exponents and equation of state
of three-dimensional spin models}

\author{M.\ Campostrini,$^{\rm a}$
        M.\ Hasenbusch,$^{\rm b}$
        A.\ Pelissetto,$^{\rm c}$
        P.\ Rossi,$^{\rm a}$ and
        E.\ Vicari$^{\rm a}$ \\
\vskip8pt
{$^{\rm a}$Dipartimento di Fisica
dell'Universit\`a di Pisa and I.N.F.N., I-56126 Pisa, Italy} \\
\vskip8pt
{$^{\rm b}$Institut f\"ur Physik,
Humboldt-Universit\"at zu Berlin, Invalidenstr.\ 110,
D-10115 Berlin, Germany} \\
\vskip8pt
{$^{\rm c}$Dipartimento di Fisica
dell'Universit\`a di Roma I and I.N.F.N., I-00185 Roma, Italy}
}
       
\begin{document}

\begin{abstract}
Three-dimensional spin models of the Ising and XY universality classes
are studied by a combination of high-temperature expansions and Monte
Carlo simulations.  Critical exponents are determined to very high
precision.  Scaling amplitude ratios are computed via the critical
equation of state.  Our results are compared with other theoretical
computations and with experiments, with special emphasis on the
$\lambda$ transition of ${}^4\rm He$.
\end{abstract}

\maketitle

\section{INTRODUCTION}

The notion of universality is central to the modern understanding of
critical phenomena.  It is therefore very important to compare
high-precision theoretical and experimental determinations of
universal quantities, such as critical exponents or universal ratios
of amplitudes, for systems belonging to the same universality class.

Critical exponents and amplitudes parametrize the singular behavior of
thermodynamical quantities in the vicinity of a critical point.
In the high-temperature (symmetric) phase $t>0$,
$C_H \approx A^+ |t|^{-\alpha}$, 
$\chi \approx C^+ |t|^{-\gamma}$,
$\xi \approx f^+ |t|^{-\nu}$, 
where $t=(T-T_c)/T_c$ is the reduced temperature, $C_H$ is the
specific heat, $\chi$ is the magnetic susceptibility, and $\xi$ is the
correlation length.
In the low-temperature (broken) phase $t<0$, $H\to0$,
$C_H \approx A^- |t|^{-\alpha}$,
$\chi \approx C^- |t|^{-\gamma}$,
$\xi \approx f^- |t|^{-\nu}$ (in the case of Ising),
$M \approx B |t|^{\beta}$,
where $M$ is the magnetization.
On the critical isotherm $t=0$, $H\ne0$,
$\chi \approx C^c |H|^{-\gamma/\beta\delta}$,
$\xi \approx f^c |H|^{-\nu/\beta\delta}$.
At the critical point $t=0$, $H=0$ at nonzero momentum,
the two-point function behaves like
$\widetilde G(q) \approx D q^{\eta-2}$.

The {\em critical exponents} are universal, and are independent of the
phase; they are related by the scaling and hyperscaling relations.
The amplitudes are not universal, and their value depends on the
phase; it is however possible to define {\em universal ratios of
amplitudes}, which are independent of the normalization of $H$, $M$,
and $T$.

The universality classes of Ising and XY in three dimensions have been
the subject of many theoretical studies.  Nonetheless, we believe that
further refinement is worthwile: many critical phenomena in nature
fall into these classes, and the precision of experiments is ever
improving; moreover, several theoretical techniques can be applied and
compared to each other.

High-temperature (HT) series expansion is one of the oldest and most
successful approaches to the study of critical phenomena.  We are
extending the length of the series available for wide families of
models belonging to the classes of universality we are interested in;
so far we computed the two-point function of the three-dimensional
Ising class to 25th order on the {\rm bcc} lattice, and four-, six-...
point functions to 21st, 19th... order \cite{LCE}.  Work is in
progress on the {\rm sc} lattice and on the XY class.

The precision of the results which can be extracted from long HT
series is mainly limited by the presence of confluent corrections with
noninteger exponents.  Let us consider, e.g., the magnetic
susceptibility $\chi$; near the critical temperature, it behaves like
\begin{eqnarray*}
\chi &=& C t^{-\gamma} \bigl( 1 + a_{0,1} t + a_{0,2}t^2 + ...\\
&&+\, a_{1,1} t^\Delta + a_{1,2} t^{2\Delta} + ... 
+ a_{2,1} t^{\Delta_2} + ... \bigr).
\end{eqnarray*}
While the exponents $\Delta$, $\Delta_2$, ... are universal, the
coefficients $a$ are model dependent. For the models we are interested
in, $\Delta\sim0.5$ and $\Delta_2\gtrsim2\Delta$; therefore it is very
helpful to select one-parameter families of models, and tune the
irrelevant parameter $\lambda$ to the special value $\lambda^\star$
for which $a_{1,1} = 0$; we will call such models ``improved''.

Monte Carlo (MC) algorithms and finite-size scaling techniques are
very effective in the determination of $\lambda^\star$ and $\beta_c$,
but not as effective in the computation of critical exponents or other
universal quantities.  On the other hand, the analysis of HT series is
very effective in computing universal quantities, but not in computing
$\lambda^\star$ and $\beta_c$.

The strength of the two methods can be combined by computing
$\lambda^\star$ and $\beta_c$ by MC, and feeding the resulting values
into the analysis of HT series (by ``biasing'' the analysis); this
greatly improves the quality of the results.

In order to keep systematic errors under control, we always select
several different families of models in the same universality class
and check that they give compatible results for universal quantities.

\section{CRITICAL EXPONENTS}

Without further discussion, we present in Table~\ref{table:1} a
selection of results for the critical exponents $\gamma$, $\nu$, and
$\eta$ of the three-dimensional Ising model; for other exponents, see
Ref.\ \cite{IHT-PRE}.  We compare the most precise theoretical results
and experiments.  IHT denotes our results \cite{IHT-PRE}; HT is a
``traditional'' HT determination \cite{B-C-97}; MC are Monte Carlo
results \cite{Hasenb-99}; FT are results from a $g$ expansion in fixed
dimension \cite{G-Z-98}.  Experimental results are LV for liquid-vapor
transitions; BM for binary mixtures; MS for uniaxial magnetic systems;
MI for micellar systems; cf.\ Refs.\ \cite{IHT-PRE} and
\cite{P-H-A-91} for bibliographical details.  The agreement between
the different determinations is overall satisfactory.

\begin{table}[t]
\caption{Comparison of determinations of critical exponents of the
three-dimensional Ising model.}
\label{table:1}
\begin{tabular}{lr@{}lr@{}lr@{}l}
\multicolumn{1}{c}{}&
\multicolumn{2}{c}{$\gamma$}&
\multicolumn{2}{c}{$\nu$}&
\multicolumn{2}{c}{$\eta$}\\
\hline
IHT & $1$&$.2371(4) $ &  $0$&$.63002(23)$ & $0$&$.0364(4) $ \\
HT  & $1$&$.2384(6) $ &  $0$&$.6308(5)  $ &    &   \\
MC  & $1$&$.2367(11)$ &  $0$&$.6296(7)  $ & $0$&$.0358(9) $ \\
FT  & $1$&$.2405(15)$ &  $0$&$.6300(15) $ & $0$&$.032(3) $ \\
\hline                                                      
LV  & $1$&$.233(10) $ &     &             & $0$&$.042(6)  $ \\
BM  & $1$&$.228(39) $ &  $0$&$.628(8)   $ & $0$&$.0300(15)$ \\
MS  & $1$&$.25(2)   $ &  $0$&$.64(1)    $ &    &   \\
MI  & $1$&$.237(7)  $ &  $0$&$.630(12)  $ & $0$&$.039(4)  $ \\
\hline             
\end{tabular}
\end{table}

On the theoretical side, similar techniques can be applied to the XY
model, with results of comparable quality.  The experimental situation
is quite different: one extremely precise experiment on the $\lambda$
transition of ${}^4\rm He$ \cite{LSNCI-96} overshadows the field.  We
present results for the critical exponents $\gamma$, $\eta$, and
$\alpha$ (we remind that $d\nu=2-\alpha$) in Table~\ref{table:2} (cf.\ 
footnote 2 in Ref.\ \cite{IHT-CHPRV} for discussion of the
experimental results).  Theoretical results are taken from Refs.\ 
\cite{IHT-PRB1} (IHT), \cite{IHT-CHPRV} (IHT${}^\star$), \cite{B-C-97}
(HT), \cite{IHT-CHPRV} (MC), and \cite{G-Z-98} (FT).  There is
disagreement between IHT${}^\star$ and experiment; it would be
interesting to improve further the theoretical computation, and to
have an independent confirmation of the experimental measurement.

\begin{table}[t]
\setlength\tabcolsep{5pt}
\caption{Comparison of determinations of critical exponents of the
three-dimensional XY model.}
\label{table:2}
\begin{tabular}{lr@{}lr@{}lr@{}l}
\multicolumn{1}{c}{}&
\multicolumn{2}{c}{$\gamma$}&
\multicolumn{2}{c}{$\eta$}&
\multicolumn{2}{c}{$\alpha$}\\
\hline
${}^4\rm He$ & & & & & $-0$&$.01056(38) $ \\
\hline             
IHT &  $1$&$.3179(11) $ & $0$&$.0381(3)$  & $-0$&$.0150(17) $ \\
IHT${}^\star$ &
       $1$&$.3177(5)  $ & $0$&$.0380(4)$  & $-0$&$.0146(8)  $ \\
HT  &  $1$&$.322(3)   $ & $0$&$.039(7)$   & $-0$&$.022(6)   $ \\
MC  &  $1$&$.3177(10) $ & $0$&$.0380(5)$  & $-0$&$.0148(15) $ \\
FT  &  $1$&$.3169(20) $ & $0$&$.0354(25)$ & $-0$&$.011(4)   $ \\
\hline
\end{tabular}
\end{table}

\section{CRITICAL EQUATION OF STATE}

The critical equation of state is a relation between thermodynamical
quantities which is valid in both phases in the neighborhood of the
critical point (cf., e.g., Ref.\ \cite{ZJbook}).

In order to determine the critical equation of state, we start from the
effective potential (Helmholtz free energy)
\[
{\cal F} (M) = M H - {1\over V} \log Z(H).
\]
In the high-temperature phase, ${\cal F}$ can be expanded in powers of
$M^2$ around $M=0$.
%
By choosing appropriately the normalizations of the renormalized
quantities, and using the ``second moment'' mass $m$ as mass scale, we
can write
\begin{eqnarray*}
\Delta{\cal F} &\equiv& {\cal F} (M) - {\cal F} (0) =
{m^d\over g_4} A(z), \\
A(z) &=& {1\over2} z^2 + {1\over4!} z^4
+ \sum_{j\ge3} {1\over(2j)!} r_{2j} z^{2j},
\end{eqnarray*}
where $z$ is the (rescaled) zero-momentum vacuum expectation value of
the renormalized field, $r_{2j}=g_{2j}/g_4^{j-1}$, and $g_{2j}$ is the
renormalized zero-momentum $2j$-point coupling constant.  The
(universal) critical limit of $g_4$ and $r_{2j}$ can be computed from
the HT expansion of the zero-momentum $2j$-point Green's function; for
the Ising model, we obtain \cite{IHT-PRE}
\begin{eqnarray*}
&g_4 = 23.49(4), \qquad
r_6 = 2.048(5), \\
&r_8 = 2.28(8), \qquad
r_{10} = -13(4).
\end{eqnarray*}

The equation of state can now be written as
\begin{eqnarray}
H(M,t) &=& {\partial{\cal F}\over\partial M} \propto
t^{\beta\delta}{\D A\over\D z} = t^{\beta\delta} F(z),
\label{eq:H} \\
\qquad z &\propto& M t^{-\beta}; \nonumber
\end{eqnarray}
The analyticity properties of $F(z)$ are constrained by
Griffiths' analyticity conditions on $H(M,t)$.

It is possible to implement all analyticity and scaling properties of
the critical equation of state introducing a parametric representation
\cite{G-Z-97} 
\begin{eqnarray*}
M &=& m_0 R^\beta \theta,\\
t &=& R(1-\theta^2), \\
H &=& h_0 R^{\beta\delta} h(\theta), 
\qquad h(\theta) = \theta + O(\theta^3).
\end{eqnarray*}
The following correspondences should be noticed:
\begin{eqnarray*}
\theta=0 &\ \longrightarrow\ & t>0, M=0; \\
\theta=1 &\ \longrightarrow\ & t=0; \\
\theta=\theta_0 &\ \longrightarrow\ & t<0, M=M_0,
\end{eqnarray*}
where $\theta_0$ is the first positive zero of $h(\theta)$.  The
analytic properties of the equation of state are reproduced if
$h(\theta)$ is analytic in the interval $[0,\theta_0]$.

Combining the parametric representation with Eq.\ (\ref{eq:H}), we
obtain
\begin{eqnarray*}
z &=& \rho\,\theta^2 (1-\theta^2)^{-\beta}, \\
h(\theta) &=& \rho^{-1} (1-\theta^2)^{\beta\delta} F\bigl(z(\theta)\bigr).
\end{eqnarray*}

In the Ising case, corresponding to the breaking of a discrete
symmetry, $\theta_0$ is a simple zero of $h(\theta)$.  We approximate
$h(\theta)$ with an odd polynomial in $\theta$, fixing its
coefficients from the small-$z$ expansion of $F(z)$.  $\rho$ is a free
parameter; as long as we keep the parametric representation exact its
value is immaterial, but it becomes significant once we make
approximations.  $\rho$ can be used to optimize the approximation, and
it can be determined from a global stationarity
condition \cite{IHT-PRE}.

We use the values of $\beta$, $\delta$, $r_6$, $r_8$, $r_{10}$
obtained by IHT to compute successive approximations to $h(\theta)$;
we check the stability of the values of several universal amplitude
ratios in order to select the best approximation.  Among the many
amplitude ratios which can be computed from $h(\theta)$, we report in
Table~\ref{table:3} $U_0=A^+/A^-$, $Q_c = B^2(f^+)^3/C^+$, $U_\xi =
f^+/f^-$; many more ratios can be found in Ref.\ \cite{IHT-PRE}.
HT+LT is a combination of HT and low-temperature expansion
\cite{L-F-89,F-Z-98}; the other theoretical determinations are the
same discussed for the critical exponents, and are taken from Refs.\ 
\cite{IHT-PRE} (IHT), \cite{C-H-97,H-P-98} (MC), and
\cite{L-M-S-D-98,B-B-M-N-87,G-K-M-96} (FT).  For experimental data,
see Refs.\ \cite{IHT-PRE} and \cite{P-H-A-91}.  The agreement between
the different determinations is again satisfactory.

\begin{table}[t]
\caption{Comparison of determinations of universal ratios of
amplitudes of the three-dimensional Ising model.}
\label{table:3}
\begin{tabular}{llll}
\multicolumn{1}{c}{ } &
\multicolumn{1}{c}{$U_0$} &
\multicolumn{1}{c}{$Q_c$} &
\multicolumn{1}{c}{$U_\xi$} \\
\hline
IHT  & $0.530(3)$  & $0.3330(10)$ & $1.961(7)$ \\
HT+LT& $0.523(9)$  & $0.324(6)$ & $1.96(1)$  \\
MC   & $0.560(10)$ & $0.328(5)$ & $1.95(2)$ \\ 
MC   & $0.550(12)$ \\
FT   & $0.540(11)$ & $0.331(9)$ & $2.013(28)$ \\
\hline
BM   & $0.56(2)$ & $0.33(5)$ & $1.93(7)$ \\
MS   & $0.51(3)$ & & $1.92(15)$ \\
LV   & $0.538(17)$ & $0.35(4)$ \\
\hline
\end{tabular}
\end{table}

In the XY case, corresponding to the breaking of a continuous
symmetry, $\theta_0$ is a double zero of $h(\theta)$.  We therefore
set 
\[
h(\theta) = \theta\bigl(1-\theta^2/\theta_0^2\bigr)^2
(1 + c_2 \theta^2 + c_4 \theta^4 ...).
\]
We fix $\theta_0$, $c_2$, ... from the small-$z$ expansion of $F(z)$,
and $\rho$ from the requirement $h(\theta)\approx(\theta_0-\theta)^2$
for $\theta\to\theta_0$.

Only the ratio $A^+/A^-$ is measured experimentally to high precision
\cite{LSNCI-96}.  We report a selection of theoretical determinations:
IHT+${}^4\rm He$ is our IHT computation, using as input for $\alpha$
the experimental value $\alpha = -0.01285(38)$
 \cite{IHT-PRB2}; IHT${}^\star$ is a
complete IHT computation, without experimental input \cite{IHT-CHPRV};
FT is a $g$ expansion in fixed dimension \cite{L-M-S-D-98};
$\varepsilon$-exp is obtained by $\varepsilon$ expansion
\cite{B-B-M-N-87}.  The value of $A^+/A^-$ is strongly correlated
with the value of $\alpha$, and all disagreement between IHT${}^\star$
and experiment can be reconduced to the discrepancy in $\alpha$.

\begin{table}[t]
\caption{Comparison of determinations of the universal ratio
$U_0=A^+/A^-$ of the three-dimensional XY model.}
\label{table:4}
\begin{tabular}{lr@{}l}
\multicolumn{1}{c}{}&
\multicolumn{2}{c}{$A^+/A^-$}\\
\hline
${}^4\rm He$      & $1$&$.0442$ \\
\hline             
IHT+${}^4\rm He$  & $1$&$.055(3)$ \\
IHT${}^\star$     & $1$&$.062(4)$ \\
FT                & $1$&$.056(4)$ \\
$\varepsilon$-exp & $1$&$.029(13)$ \\
\hline
\end{tabular}
\end{table}

\section{CONCLUSIONS}

The study of HT series of ``improved'' models, with parameters
determined by MC simulations, allowed us to compute with high
precision the universal quantities (critical exponents and effective
potential) characterizing the critical behavior of the symmetric
phase.

Suitable approximation schemes allow the reconstruction of the
critical equation of state starting from the symmetric phase; many
universal amplitude ratios can be computed.

For the Ising universality class, theoretical computations are much
more precise than experiments.  On the other hand, for the XY class,
some very precise experimental results for $\alpha$ and $A^+/A^-$ have
been obtained \cite{LSNCI-96}.  There is disagreement with the most
precise theoretical results \cite{IHT-CHPRV}.  A new-generation
experiment is in preparation \cite{Nissen-98}; it would be interesting
to improve further the theoretical computations as well.

\end{document}